\documentstyle[12pt]{article}

\newcommand{\nc}{\newcommand}
\nc{\beq}{\begin{equation}}
\nc{\eeq}{\end{equation}}
\nc{\beqa}{\begin{eqnarray}}
\nc{\eeqa}{\end{eqnarray}}

\input epsf
\newwrite\ffile\global\newcount\figno \global\figno=1

\def\writedef#1{}

\def\figin{\epsfcheck\figin}\def\figins{\epsfcheck\figins}
\def\epsfcheck{\ifx\epsfbox\UnDeFiNeD
\message{(NO epsf.tex, FIGURES WILL BE IGNORED)}
\gdef\figin##1{\vskip2in}\gdef\figins##1{\hskip.5in}
instead
\else\message{(FIGURES WILL BE INCLUDED)}%
\gdef\figin##1{##1}\gdef\figins##1{##1}\fi}

\def\figinsert{}
\def\ifig#1#2#3{\xdef#1{fig.~\the\figno}
\writedef{#1\leftbracket fig.\noexpand~\the\figno}%
\figinsert\figin{\centerline{#3}}\medskip\centerline{\vbox{\baselineskip12pt
\advance\hsize by -1truein\center\footnotesize{  Fig.~\the\figno.} #2}}
\bigskip\endinsert\global\advance\figno by1}
\def\endinsert{}

\begin{document}

\title{\large{\bf Fractional instantons\\
 in supersymmetric gauge theories}}

\author{\\Maxim Zabzine\thanks{zabzin@physto.se}\\
\textit{Institute of Theoretical Physics}\\
\textit{University of Stockholm}\\
\textit{Box 6730}\\
\textit{S-113 85  Stockholm, Sweden}}

\date{}

\maketitle

\begin{picture}(0,0)(0,0)
\put(390,295){USITP-98-15}
\put(390,280){hep-th/9807120}
\end{picture}
\vspace{-24pt}

\begin{abstract}
We consider evidence for the existence of
  gauge configurations with fractional charge in 
 pure $N=1$ supersymmetric Yang-Mills theory . We argue that
 these field configurations are singular and have to
 be treated as distributions. It is shown that 
 the path integral  representation of constant Green's functions
 can be reduced to a finite dimensional integral. The 
 fractional configurations are 
 essentially the zero size limit of the usual instantons and they have 
 a reduced number of fermionic zero modes.     
At the end we comment on the status of 
  the $D$-instanton/YM-instanton correspondence within 
 $AdS$/CFT correspondence.  
\end{abstract}

\newpage
\section{Introduction}

In the last few years a number of remarkable results 
 on the non-perturbative behavior in gauge theories and
 string theories have appeared. The exact solutions
 for supersymmetric gauge theories are striking in their 
 self-consistency and seem to refute any doubts in
 their validity. One of the basic result is the 
 existence of a non-zero gluino condensate in pure $N=1$ supersymmetric
 Yang-Mills theory. 
 There are different consistent approaches to get
 non-zero gluino condensate, from the Veneziano-Yankielowicz
  effective action \cite{vy}
 to the instanton calculations \cite{instanton}. Also the existence
 of a non-zero gluino condensate is consistenly incorporated 
 in the full picture of supersymmetric gauge and string theories.
 Nevetheless within the framework of QFT there is still 
 a confusing  question: What kind of field configurations
 gives rise the gluino condensate itself? 
 The solution of this problem might be relevant to the
 understanding of confinment in the supersymmetric gauge
 theories. There has been 
 an attempt to answer this question in a finite volume
 \cite{gomez} by introducing twisted boundary conditions.
 To the author's knowledge the only attempt
 to answer this question in infinite volume is by introducing
  multivalued field configurations \cite{zhit}. In the
 present letter we try to give an answer for the 
 problem 
  without any direct references to the ideas of \cite{gomez} 
 and \cite{zhit}.

The main puzzle is that the existence of a non-zero gluino 
 condensate requires gauge configurations
 with fractional topological charge . This goes 
 beyond our present understanding of QFT. There are no
  smooth gauge field configuration with finite action 
 which can have a
 reduced number of fermionic zero modes. According to standard
 wisdom, all gauge configurations are classified by 
 integer topological charge and the path integral is 
 just a sum over different topological sectors. The above 
 problem can be resolved if one is willing to study non smooth 
 configurations which nevertheless have finite 
 action\footnote{In the sense that one has a unique recipe to 
 define the action for such configurations.}.
 We are going to look for singular 
 configurations with the right properties. We argue that
 in fact the best candidate for this role is a configuration with
 point support which can be regarded as zero-size
 instanton.  We think that
 it is an important lesson for gauge theories to understand
 that perhaps the path integral is not just a sum over 
 different topological sectors of smooth configurations
  but also a sum over some 
 singular configurations. 

The organization of the paper is as follows. In section
 2 we review the basic facts about pure supersymmetric
 Yang-Mills theories. We recall the chiral selection
 rules and show that non-zero gluino condensate 
 requires a gauge configuration with two fermionic zero 
 modes. In section 3 we consider a QFT with constant 
 Green's functions (SYM is example of such theory). Using
 these constant Green's functions we argue  
 that  the configurations with point
 support are important in the path integral and that only
 the topological part survives 
 in the action (therefore it is just a number).
 In section 4 we consider the configurations with
 point support  and show that in fact they have the right 
 number of fermionic zero modes. We point out that the zero-size
 instantons are essentially merons, configurations which were
 inroduced by hand in \cite{gross} to get the area law for
 Wilson loop operator.
 In section 5 we give 
 our conclusions and comment on the status of 
  the $D$-instanton/YM-instanton correspondence within 
 $AdS$/CFT correspondence.

\section{Evidence for the fractional configurations}

First of all let us recall the basic relevant facts 
 about the pure supersymmetric gauge theory in four dimensions. 
The SYM Lagrangian describing the gluodynamics of gluons $A_\mu$ and 
gluinos $\lambda_\alpha$ with a general compact gauge group has the form 
\beq
\label{a1}
{\cal L} = -\frac{1}{4g_0^2} 
G_{\mu\nu}^a G_{\mu\nu}^a
+ \frac{i}{g_0^2}\lambda_{\dot\alpha}^\dagger 
D^{\dot\alpha\beta}\lambda_\beta
+ \frac{i\theta}{32\pi^2} G_{\mu\nu}^a \tilde{G}_{\mu\nu}^a. 
\eeq
where $G_{\mu\nu}^a$ is the gluon field strength tensor, 
$\tilde{G}_{\mu\nu}^a$
is the dual tensor and $D^{\dot\alpha \beta}$ is the covariant
derivative and all quantities are defined with respect to the adjoint 
representation of the gauge group. This  Lagrangian may be written
in terms of  the  gauge superfield $W_\alpha$
with physical components ($\lambda_\alpha$, $A_\mu$) as follows
\beq
\label{a2}
{\cal L} = \frac{1}{8\pi} Im \int d^2\theta\tau_0 W^{\alpha}W_{\alpha}, 
\eeq
where the bare gauge coupling $\tau_0$ is defined to be
$\tau_0 = \frac{4\pi i}{g^2_0} + \frac{\theta_0}{2\pi}$.
The model possesses a discrete global $Z_{2C_2}$ symmetry\footnote{$C_2$
denotes the Dynkin index (the quadratic Casimir) with $C_2 = 1/2$
 normalization for the fundamental 
 representations of $SU(N)$, $Sp(N)$ and with $C_2=1$ 
normalization for
 the vector representation of $SO(N)$. So for adjoint 
representations we use
 the following values: $C_2(SU(N))=N$, $C_2(Sp(2N))=N+1$,
 $C_2(SO(N))=N-2$,
 $C_2(E_6)=12$, $C_2(E_7)=18$, $C_2(E_8)=30$, $C_2(F_4)=9$,
 $C_2(G_2)=4$. }, a residual 
non-anomalous subgroup of the anomalous chiral $U(1)$.
 The discrete chiral symmetry $Z_{2C_2}$ is sponteneously broken
 by a non-zero gluino condensate $\langle \lambda^{a\alpha}
 \lambda^a_\alpha \rangle \equiv \langle \lambda \lambda\rangle$.
 In term of the strong coupling scale $\Lambda$ the gluino
 condesate has the form
\beq
\label{a3}
\langle \lambda \lambda \rangle = c\, \Lambda^3\,
e^{\frac{2\pi i k}{C_2}},\,\,\,\,\,k\in Z.
\eeq
where the constant $c$ can be absorbed in a redefinition
 of $\Lambda$.
Therefore the system has $C_2$ different values of
 the gluino condensate and can be in any one of 
 the $C_2$ vacua. In what follows we suppose that the system
 sits in one of the vacua and that all calculations are valid 
 for this phase. The supersymmetry requires that 
 the Green's functions of the composite operator $\lambda\lambda$
 do not depend on the coordinates. Due to the cluster 
 decomposition property we have the following equalities
\beq
\label{a4}
\langle \lambda\lambda(x_1) \lambda\lambda(x_2) ...
 \lambda\lambda(x_n)\rangle = \Lambda^{3n},\,\,\,\,
\langle \lambda\lambda(x) \rangle = \Lambda^3.
\eeq
We mainly concentrate our attention on SYM
 with an $SU(2)$ gauge group.  

 Now we consider the arguments 
 supporting the existence of fractional topological
 charge. It is beleived that the path intergral is a sum
 over different topological sectors. Let us look at 
 the Green's function restricted to one topological sector. 
 Consider a chiral rotation of fermions
 $\lambda \rightarrow e^{i\alpha} \lambda$ which 
 gives the following rotation in the path integral
 representation of the relevant Green's functions
\beq
\label{a5}
\langle \lambda\lambda(x_1) \lambda\lambda(x_2) ...
 \lambda\lambda(x_n)\rangle = e^{(2in\alpha - i\alpha f)}
 \int DA\, D\lambda\,\,\, \lambda\lambda(x_1) \lambda\lambda(x_2) ...
 \lambda\lambda(x_n) e^{-S(A,\lambda)}
\eeq
where $f=\nu_{+}-\nu_{-}$  and $\nu_{+}$ ($\nu_{-}$)
 is number of fermionic zero modes with positive (negative)
 chirality. The equality (\ref{a5}) is just a change of variables
 in path integral. The exponent in front of the path integral comes
 from an explicit rotation of the fermion fields and the anomalous
 rotation of the mesure.
 The last contribution can be derived using the Fujikawa's method 
 \cite{fujikawa}. Since the expression is  independent
 of $\alpha$ we can take derivative with respect to $\alpha$ 
and get the following
 chiral selection rules
\beq
\label{a6}
(2 n - f)\langle \lambda\lambda(x_1) \lambda\lambda(x_2) ...
 \lambda\lambda(x_n)\rangle = 0.
\eeq
In fact the number fermionic zero modes is related to
 the properties of the gauge field. We are interested in the 
 continuous gauge configurations which give a finite action. Therefore
 all such continuous configurations are characterized by
 the third homotopy group $\pi_3(G)$. The corresponding topological
 charge is given by the following expression
\beq
\label{a7}
Q= \frac{1}{32\pi^2} \int d^4 x\,Tr(\epsilon^{\mu\nu\rho\sigma}
G_{\mu\nu}^a G_{\rho\sigma}^a)
\eeq
which is called the second Chern or first Pontrjagin number, depending
 on the group. The Atiyah-Singer theorem gives the dependence
 between the number of zero modes of the Dirac operator 
 in the gauge backgound and the second Chern number of this backgound
 $2C_2 Q = f$. After all one can rewrite the expression 
 (\ref{a6}) as follows
\beq
\label{a8}
(n-C_2 Q)\langle \lambda\lambda(x_1) \lambda\lambda(x_2) ...
 \lambda\lambda(x_n)\rangle = 0.
\eeq 
Therefore one concludes that that the n-point function is 
 saturated by the gauge configurations with topological number
 $Q=n/C_2$. Since we consider fermions in the adjoint 
 represetation $C_2$ is quite ``big''. The non-vanishing
 gluino condensate has $Q=1/C_2$, but $Q$ is supposed to 
 be integer. In the work on instantons \cite{instanton}, to 
 avoid contradiction, the following average over degenerate vacua
 was taken
\beq
\label{a9}
\sum\limits_{k=1}^{C_2} \langle \lambda\lambda(x_1) \lambda\lambda(x_2) ...
 \lambda\lambda(x_n)\rangle = 0, \,\,\,\,\,\,\,\,\frac{n}{C_2}\not\in Z
\eeq 
where $k$ labels the different vacua. So these averaged Green's
 functions vanish when $Q$ is  fractional. From general
 principeles of QFT we have no right to average  Green's 
 functions over different degenerate vacua. The worlds with
 different gluino condensate live completely independent lifes
 (with no correlations between them). Thus as soon as we 
 accept the
 existence of a non-vanishing gluino condensate and an unbroken
 supersymmetry we have to accept the existence of all n-point 
 Green's funtions (\ref{a4}) for the operator $\lambda\lambda$. 
 We still consider a system which sits in one of 
 the vacua.

To resolve this problem we have to assume that the gauge 
 configurations which give rise the Green's functions are
 not continous (and as a result not smooth). If we accept this 
 then the topological charge cannot classify these 
 configurations by definition. To be able to work with such
 configurations one has to go to the notion of generalised 
 function (distribution). It is an essential starting point 
 for the axiomatic approach to QFT to think about fields
 as operator valued distributions. It can be shown 
 that considering fields as ordinary functions (objects 
 defined at each point of space time) leads to  
contradictions in QFT or triviality of QFT \cite{logunov}.       
Further we will argue that the role of fractional instantons
 may be played by field configurations with point support. We
  have to understand that such configurations are not instantons 
 in the usual sense, they are singular and any reasonable calculation
 requires  regularization of these configurations. 
Nevertheless these configurations
 satisfy the self-duality conditions (in a generalized sense)
 and we may view them as instantons, at least  formally.

\section{Constant Green's functions}

In this section we are going to use the property that
 the relevant Green's funtions are constant.
 This property is a result of unbroken supersymmetry.
 To simplify the notation we will consider an abstract
 QFT and identify the composite operator $\lambda\lambda$
 with the fundamental field. All arguments can
 be strightforwardly specified to a situation with
 certain  field content.    
 Let us consider a QFT with some opeartor $\phi(x)$
 which has a constant non vanishing n-point Green's function
 $\langle \phi(x_1) \phi(x_2)...\phi(x_n)\rangle$. Due 
 to the cluster decomposition property we have the following 
 equalities
\beq
\label{b1}
\langle \phi(x_1) \phi(x_2)...\phi(x_n)\rangle = A^n,\,\,\,
\,\,\,\langle \phi(x) \rangle =A.
\eeq
The generating functional for the Green functions is
\beq
\label{b2}
Z[J(x)] = \int D\phi\,\,e^{-S(\phi) + \int d^4x\,\phi(x)J(x)},\,\,\,
\,\,\,\,Z[0]=1
\eeq
which one can expand in the standard way
\beq
\label{b3}
Z[J(x)] = \sum\limits_{n=0}^{\infty}\frac{1}{n!} \int\int ...\int
 d^4x_1\, d^4x_2...d^4x_n
J(x_1)J(x_2)...J(x_n) \langle \phi(x_1) \phi(x_2)...\phi(x_n)\rangle
\eeq
Using the equlities (\ref{b1}) we can rewrite the
 expression (\ref{b3}) as follows
\beq
\label{b4}
Z[J(x)] = \sum\limits_{n=0}^{\infty}\frac{A^n}{n!} \int  d^4x_1
J(x_1) \int  d^4x_2 J(x_2) ... \int d^4x_n J(x_n) = e^{A \int  d^4x J(x)}
\eeq
 In general it is very difficult to treat rigirously the path
 integral representation of the generating functional since this 
 object is non-linear functional of  the source. But the 
 expression (\ref{b4}) is somewhat particular. Usually one defines
 such constructions on the appropriate Banach space. Let us consider
 the generating functional on the space ${\cal L}_1(R^4)$ (the Banach
 space with norm $\|J\|_{{\cal L}_1} = \int d^4x |J(x)| < \infty$).
 The key observation is that the logarithm of the generating functional 
 (\ref{b4}) is a linear continious functional on  ${\cal L}_1(R^4)$
 ($|\log Z[J(x)]| \leq A \|J\|_{{\cal L}_1}$). Therefore we can define
 the generating functional on some dense subset of the space and
 afterwards get a unique continuation to the whole space .
We will consider sources which have the following special
 form
\beq
\label{b5}
J(x) = \left\{ \begin{array}{l}
              J=const,\,\,\,\,\,\,x\in {\cal B};\\
              0,\,\,\,\,\,\,\,x\not\in {\cal B}.
                \end{array} \right.
\eeq
where ${\cal B}$ is a set of finite volume $V$ in $R^4$ which 
is homeomorphic to the ball. The set of linear combinations of 
 the functions (\ref{b5}) is a dense subset in ${\cal L}_1(R^4)$.
 Hence for every function $J(x) \in {\cal L}_1(R^4)$ one can find
 sequence $J_n (x)$ which is a finite linear combination of
 sources $J_i$ defined in (\ref{b5})
\beq
\label{b5a}
J(x) = \lim\limits_{n} \sum\limits_i J_i(x)
\eeq
where any numbers are included in the definitions of $J_I(x)$. The
 limit in (\ref{b5a}) is understood with respect to topology given
 by $\|.\|_{{\cal L}_1}$. Thus we can write the following definition 
 of the generating functional on arbitrary function from ${\cal L}_1$
\beq
\label{b5b}
\log Z[J(x)] = \lim\limits_n \sum\limits_i (\log Z[J_i(x)])
\eeq 
 We conclude that it is enough to consider the
 generating functional restricted on the subset of sources with form
 (\ref{b5}). 
Let us concentrate our attention on the one-point function.
 The integral
 over the source (\ref{b5}) is $\int d^4x J(x) = J V$.
  We see that the functional
 $Z[J, V] = e^{AJ V}$ is essentially a function of $J$ and $V$. 
 Thus one has the following definition of the relevant Green's functions
\beq
\label{b6}
\langle \phi(x) \rangle =
\frac{1}{V}
 \left. \frac{\partial Z}{\partial J}\right|_{J=0} =
 \frac{1}{V}
 \left. \frac{\partial Z}{\partial J}\right|_{V=0} =
 \frac{1}{J}\left. 
\frac{\partial Z}{\partial V}\right|_{V=0}.
\eeq
These definitions of constant Green's functions are somewhat
 suprising since, after taking derivatives, one has to take the 
 zero volume limit. Let us look more carefully at the construction
\beq
\label{b7}
Z[J, V] = \int D\phi\,e^{-S(\phi) + J 
\int\limits_{\cal B} d^4x\, \phi(x)} 
\eeq
where the action is an integral over all of space-time 
 $S(\phi) = \int\limits_{R^4} d^4x\, {\cal L}(\phi)$. Every function
 $\phi(x)$ can be decomposed as $\phi(x)= \phi_1(x) + \phi_2(x)$
 where $\phi_1(x)$ has support inside the ball ${\cal B}$
 and $\phi_2(x)$ has support
 outside ball ${\cal B}$. In the same way we can decompose
 the action $S(\phi)=S_1(\phi)+S_2(\phi)$ where the first part is
 integral over the ball and second is integral over its complement. 
 In this separation of the action the values of $\phi(x)$ 
 on the boundary of the ball are   important 
 since non-trivial boundary conditions
 can give rise important surface contributions. We assume that
 all possible surface contributions are included in $S_1$.
 Therefore the generating functional can be written in the following
 form
\beq
\label{b8}
Z[J,V]=\frac{\int D\phi_2\, e^{-S_2(\phi_2)}\,\int D\phi_1\,
e^{-S_1(\phi_1) + J \int\limits_{\cal B} d^4x\, \phi(x)}}
{\int  D\phi_2\, e^{-S_2(\phi_2)}\,\int D\phi_1\,
e^{-S_1(\phi_1)}}
\eeq   
where in the denominator we have written out the contribution 
$Z[J(x)=0]$  which we normalized to one in (\ref{b2}).    
In (\ref{b8}) we did not write explicitly the integration 
 over collective coordinates. The ball ${\cal B}$ has a 
 center $x_0$ and one has to integrate out this dependence
 in the expression (\ref{b8}). In the supersymmetric models 
 we can put this ball in superspace and thus we have two 
 collective coordinates $x_0$ and its superpartner $\theta_0$.
 Hence we can write the generating functional as follows
\beq
\label{b9}
Z[J,V]=\int d^4x_0\,\int D\phi\,
e^{-\int\limits_{\cal B} d^4x {\cal L}  + 
 J \int\limits_{\cal B} d^4x\, \phi(x)}
\eeq
where in general the integration over appropriate collective
 coordinates are assumed. Now let us use the definition (\ref{b6})
 of the Green's functions
\beq
\label{b10}
\langle \phi(x) \rangle =
 \frac{1}{V}\left.\frac{\partial Z}{\partial J}\right|_{V=0}=
 \int d^4x_0\, \int D\phi\,
\left( \frac{\phi}{V}
 \right) \left.e^{- \int\limits_{\cal B} d^4x {\cal L}}
 \right|_{V\rightarrow 0}.
\eeq
 where $\phi \equiv \int\limits_{\cal B} d^4x\, \phi(x)$. We
 have to take the zero volume limit  in the expression (\ref{b10}). In 
 this limit in the action only topological contributions survive
 which are independent of the volume $S(\phi) \rightarrow S_{top}$.
 In the zero volume limit only the configurations with point support
\footnote{In the expression (\ref{b11}) we do not specify the 
 delta function
 (for example, $\delta^4(x)$ or $\delta(x^2)$). It may depend of
 the details of the limiting procedure.} 
 survive 
\beq
\label{b11}
\frac{\phi}{V} \rightarrow \phi\delta(x-x_0)\,\,\,
\Longleftrightarrow\,\,\,\int d^4x \frac{1}{V}
 \phi(x) \rightarrow \int d^4x \delta(x-x_0)\phi(x),\,\,\,\,\,\,\,\,\,
V\rightarrow 0.
\eeq  
These configurations give rise to constant Green's functions. 
 Therefore schematiclly we can write down the following expression
 for the Green's functions
\beq
\label{b12}
\langle \phi(x) \rangle = \sum\limits_{top}
 \int d^4x_0\,\int d\phi\, \phi\, \left(\delta(x-x_0)\right)\, 
 e^{-S_{top}}
\eeq
where $\int d\phi\,\phi$ is just the finite dimensional integral
 over the relevant moduli. Also the sum over different topological
 contributions has to be assumed in the most general situation.
 At this point we have to issue 
 a warrning. We cannot regard
  the expression (\ref{b12}) as a simple recipe for calculating
 the relevant Green's functions. The configurations with point
 support is UV singular and any realiable calculation will need
 an UV regularization. In the theory of distributions one cannot make
  sense of products of delta functions except by regularization.
 Aslo one has to
 worry about the proper definition the topological contribution $S_{top}$
 in the action for singular configurations. As we will see  in the
 case of Green's functions of the composite operator $\lambda\lambda$,
 one can define the topological contribution using the 
 Atiyah-Singer theorem. 

In this section we have tried to give 
 arguments in favour of the
 rather intuitivly obvious idea that the constant Green's
 functions are saturated by the configurations 
 localized at a point\footnote{This property
 is related to the fact that the relevant Green's functions
 are a set of disconnected coordinate independent pieces.}. 
 We have shown that only the configurations
 with non trivial topology (the action on these configurations is
  proportional the topological charge) can give rise the constant 
 Green's functions. 
 The space of such configurations
 is finite dimensional, therefore the path integral 
 represenation of the relevant Green's functions is reduced 
 to a finite dimensional integral over the corresponding 
 collective coordinates. In the next section we will see that
 the configurations with point support have the right number
 of fermionic zero modes.

\section{Fractional instantons}

In the last section we have argued that the field configurations with
 point support give rise constant Green's functions. Now 
 let us turn to the discussion of the situation for $SU(2)$ SYM.
 As we have seen in Section 2, for saturation
 of the gluino condensate one needs two zero fermionic modes and  
 a gauge configuration with $Q=1/2$. Let us look at the fermionic
 zero modes for self-dual (antiself-dual) gauge configurations.
 A supersymmetry transformation applied to the bosonic solution
 $G^a_{\mu\nu}$ generates two fermionic zero modes
\beq
\label{c1}
\lambda^{a}_{\alpha} \sim G^{a}_{\alpha\beta} \epsilon^{\beta}
\eeq
where $\epsilon^{\beta}$ is the spinor parameter of
 transformation. The remaining two zero modes can be obtained by
 applying a superconformal transformation to the bosonic
 solution
\beq
\label{c2}
 \lambda^{a}_{\alpha} \sim G^{a}_{\alpha\beta}(x-x_0)^{\beta
 \dot\gamma} \xi_{\dot\gamma}
\eeq
where $(x-x_0)^{\beta\dot\gamma} \xi_{\dot\gamma}$ is coordinate
 dependent parameter of transformation.
 The first two zero modes (\ref{c1}) corespond to the physical symmetry 
 of the QFT.
 The existence of the second two modes (\ref{c2}) corespond to the
 symmetry of the classical equations of motion. Therefore we
 can sacrify the superconformal modes.  
We can try to solve the problem by brute force,
  requiering the superconformal modes to vanish
\beq
\label{c3}
 G^{a}_{\alpha\beta}(x-x_0)^{\beta
 \dot\gamma} \xi_{\dot\gamma} = 0.
\eeq
This means that the superconformal transformation acts trivially on 
 the bosonic solution $ G^{a}_{\alpha\beta}$.
 Equation (\ref{c3}) has only one  solution in terms of 
 distributions which due to the dimension of the field
 can be written as follows
$G \sim \delta((x-x_0)^2)$. We have thus found that the field 
 configuration with point support has presicely two fermionic
 zero modes as we need for non-zero gluino condensate. 

The solution found above can be regarded as the zero size
 limit of the one instanton solution
\beq
\label{c4}
G^{a}_{\mu\nu} = -4\eta^{a}_{\mu\nu}\frac{\rho^2}{[(x-x_0)^2
+\rho^2]^2}\,\,\,\,\,\,\,\, \stackrel{\rho \rightarrow 0}{\longrightarrow}
\,\,\,\,\,\,\,\, -4\eta^{a}_{\mu\nu} \delta((x-x_0)^2).
\eeq
where $\eta^{a}_{\mu\nu}$ are the 't Hooft symbols.
The $-4\eta^{a}_{\mu\nu} \delta((x-x_0)^2)$ is the generalized solution
 of the self-duality equation. Therefore formally one can write the 
 expression for the action as follows $S= 8\pi^2 Q/g^2$ where $Q$ is 
 defined by (\ref{a7}). In the case of zero size instanton we cannot
 calculate the action or topological charge directly since there are
 singularities (in general expression like 
 $\delta(x)\delta(x)$ do not make 
 sense)\footnote{Using the delta-like sequences
 one can show that $\delta(x)\delta(x)\rightarrow\delta(x^2)$.}.
 The most natural way is to use the Atiyah-Singer theorem as a
 definition of the topological charge and the corresponding action. 
 It is important to note that zero size instanton is an independent 
 object in the sense that the standard instanton is not the only
  possible regularization of the zero size instanton. There are
 infinitely many such regularizations. The consideration in 
 \cite{zhit} supports the point that such singular configurations
 exist independently. 

Another very interesting point is that if we consider the gauge
 potential for the instanton in the zero size limit
\beq
\label{c4a}
A^a_{\mu} = 2\eta^{a}_{\mu\nu} \frac{(x-x_0)_\nu}{(x-x_0)^2 + \rho^2}
\,\,\,\,\,\,\,\, \stackrel{\rho \rightarrow 0}{\longrightarrow}
\,\,\,\,\,\,\,\, 2\eta^{a}_{\mu\nu} \frac{(x-x_0)_\nu}{(x-x_0)^2}
\eeq
then one can recognize in this expression the meron configuration 
 intorduced in \cite{gross}. There are two remarkable facts about 
 this configuration. First in the  meron background the Wilson loop
 has area law and therefore there is confinment. Second the meron
 is a point like defect since it can be gauge away everywhere exept
 one point $x_0$. 

The supersymmetry transformation applied to the zero size instanton
 generate two fermionic modes 
\beq
\label{c5}
\lambda^{a}_{\alpha} \sim -4\eta^{a}_{\alpha\beta}\epsilon^{\beta}
 \delta((x-x_0)^2).
\eeq
which correspond to two collective coordinate, the center of
 zero size instanton $x_0$ and its superpartner $\theta_0$. 
 The supeconformal transformations act trivially and give zero.
 This is natural since the size of the instanton $\rho$ and 
 its superpartner are lacking. In a superspace language 
 the relevant superfield
 configuration has the form
\beq
\label{c6}
W^{\alpha}W_{\alpha} \sim \delta((x-x_0)^4) (\theta -\theta_0)^2.
\eeq 
Further one has to integrate over the collective coordinate to get 
 the gluino condensate 
\beq
\label{c7}
\langle W^{\alpha}W_{\alpha} \rangle \sim M^3\int 
d^4x_0\,d^2\theta_0\,\,
\delta((x-x_0)^4) (\theta -\theta_0)^2 e^{-\frac{4\pi^2}{g^2}}
\eeq
where $M$ is a parameter with dimension of mass which is needed
 to keep the  dimensions right. After these  calculations
 we get the gluino condensate
\beq
\label{c8}
\langle \lambda\lambda\rangle \sim c M^3  e^{-\frac{4\pi^2}{g^2}} 
\sim c\Lambda^3
\eeq
where $c$ is numerical constant and $\Lambda$ is the strong coupling scale.
 Of course the above calculations are naive since we did not talk
 about the regularization which we need to consider for the product of
 two delta functions $\delta(x)\delta(x)$. Thus we did not worry 
 about numerical factors. 

In this context the standard instanton's calculations \cite{instanton}
 can be thought of as a way of introducing a regularization. 
 The size of instanton plays the role of UV-cut-off. The 
 additional fermionic modes have to be introduced to keep
 supersymmetry unbroken within these calculations.

\section{Discussions and conclutions}

In the present letter we have tried to answer  the question of what kind
 of field configurations give rise the gluino condensate. 
We have addressed  this problem within the QFT language.
 The existence of non-zero gluino condensate and unbroken
 supersymmetry gives us two puzzles.
 The first is that, from general principles, the non-vanishing matrix 
 element $\langle \lambda\lambda\rangle$ 
 requieres the existence of fractional instantons or more 
 presicely; field configurations with a reduced number of ferminic
 zero modes. Second, a non-vanishing gluino condensate 
 gives rise a tower of constant Green's functions for the composite
 operator $\lambda\lambda$. In fact  both puzzles have the
 same answer. We have shown that the constant Green's functions  
 can be saturated by field configurations with point
 support. At the same time such configurations can be thought of
 as the zero size instanton and they have exactly right number of
 fermionic zero modes. 
 
Also within this context it is interesting to note  
 the correspondence between $D$-instantons of Type IIB
 superstring on $AdS_5\times S^5$ and YM instantons in 
 SYM living on the boundary of $AdS_5$ \cite{kogan}.  
 Within this correspondence the size of instanton
 play the role of UV cut-off (distance of $D$-istanton
 to the boundary of  $AdS_5$) and instanton itself
 is an object with point support on the boundary. 
 In the large N-limit ('t Hooft limit) the instantons do
 not survive and only fractional configurations with
  action $S \sim 1/N$ can survive. Therefore it is 
 natural to have configuration with point support (thus
 with reduced number of zero modes) on the boundary.
 Hence one can call this correspondence - $D$-instanton/fractional
 instanton correspondence.   
  We do not
 want to speculate about this subject except to point 
 out the possible parallels between the different ideas.

It would be iteresting to go further on this subject. 
 For example, it would be nice to gain a better 
 understanding of how to calculate
 different condensates in supersymmetric gauge theories 
 using   configurations with point support without 
 direct use of instantons. Another problem for further
 research could be the undestanding of the twenty years
 old calculations of Wilson loop \cite{gross} in the light
 of the presented arguments.

\begin{flushleft} {\Large\bf Acknowledgments} \end{flushleft}

\noindent The author is grateful to Hans Hansson, Ulf Lindstr\"om
 and Myck Schwetz
for useful discussions and comments.
The author was supported by the grants of the Royal Swedish Academy of
Sciences and the Swedish Institute.


\begin{flushleft} {\Large\bf Note added} \end{flushleft}

After completing this work we learned about the work \cite{brodie}
 by J.Brodie where similar ideas were discussed from the string theory 
 point of view. Also we learned about some evidence \cite{nar} 
within  lattice  calculations in favor
 of the fractional instantons.


\end{document}